\documentstyle[12pt]{article}
\font\tenrm=cmr10

\def\e{{\rm e}}
\newcommand{\be}{\begin{equation}}
\newcommand{\ee}{\end{equation}}
\newcommand{\bea}{\begin{eqnarray}}
\newcommand{\eea}{\end{eqnarray}}
\newcommand{\nn}{\nonumber}
\begin{document}
% new macro for bibliography
\renewenvironment{thebibliography}[1]
  { \begin{list}{\arabic{enumi}.}
    {\usecounter{enumi} \setlength{\parsep}{0pt}
     \setlength{\itemsep}{3pt} \settowidth{\labelwidth}{#1.}
     \sloppy
    }}{\end{list}}

\parindent=1.5pc

\begin{titlepage}
\rightline{hep-th/9307159}
\rightline{MPI-Ph/93-55}
\rightline{July 1993  }

\begin{center}{{\bf DILATONIC SPHALERONS \\
               \vglue 10pt
               AND\\
               \vglue 10pt
                 NON-ABELIAN BLACK HOLES
 \footnote{Talk given at the 15th annual MRST meeting on High
Energy Physics, Syracuse University, NY, May 14-15, 1993.
To be published in the Proceedings}
               }\\
\vglue 5pt
\vglue 1.0cm
{GEORGE LAVRELASHVILI
 \footnote{On leave of absence from Tbilisi
Mathematical Institute, 380093 Tbilisi, Georgia}
 }\\
\baselineskip=14pt
{\it Max-Planck-Institut f\"ur Physik, Werner-Heisenberg-Institut}\\
{\it F\"ohringer  Ring 6, 80805 Munich, Germany}\\
\vglue 0.3cm
{and}\\
\vglue 0.3cm
{DIETER MAISON}\\
{\it Max-Planck-Institut f\"ur Physik, Werner-Heisenberg-Institut}\\
{\it F\"ohringer Ring 6, 80805 Munich, Germany}\\
\vglue 0.8cm
{ABSTRACT}}
\end{center}
\vglue 0.3cm
{\rightskip=3pc
 \leftskip=3pc
 \tenrm\baselineskip=12pt
\noindent
We discuss properties and interpretation of recently
found globally regular and black hole solutions
of a Einstein-Yang-Mills-dilaton theory.
\vglue 0.8cm}
\end{titlepage}
{\bf\noindent 1. Introduction}
\vglue 0.4cm
\baselineskip=14pt
\noindent
Soon after Bartnik and McKinnon's discovery of smooth,
static solutions of Einstein-Yang-Mills (EYM) theory \cite{BM},
black hole solutions  of the same system where found \cite{BH}.
Stability analysis showed that all those solution are unstable
in linear perturbation theory \cite{STAB}.
It was understood that the Bartnik-McKinnon solutions are a kind
of gravitational sphalerons \cite{GRSPH}.
All these solutions are not known in closed form and were
obtained by numerical integration.
In the meantime mathematically rigorous
existence proof for all these solutions were given
\cite{YAU1},\cite{YAU2},\cite{BFM}.

On the other hand,
various theoretical considerations (e.g. String theory,
Kaluza-Klein theories, cosmological models) suggest to
supplement the gravitational field by a massless scalar
companion, a so-called  `dilaton' - a neutral scalar
field with conformal coupling to matter.
So, it is a natural step to extend the EYM theory to a
Einstein-Yang-Mills-dilaton (EYMD) theory.
This extended theory was recently studied in
 \cite{LM1}-\cite{Biz2}.

In the present paper we will discuss properties and interpretation
of globally regular and black hole solutions of such a
EYMD theory.

We will consider a EYMD theory defined by the action
\be
S=
\frac{1}{4\pi}\int\Bigl(-\frac{1}{4 G}R+
\frac{1}{2}(\partial\varphi)^2-
 {\e^{2\kappa\varphi}\over 4g^2}F^2\Bigr)\sqrt{-g}d^4x \label{act}
\ee
where $\kappa$ resp.\ $g$ denote the dilatonic resp.\ gauge
coupling constant and $G$ is Newton's constant.

This theory depends on a dimensionless parameter
$\gamma=\frac{\kappa}{\sqrt{G}}$.
In the limit $\gamma\to 0$ one gets the EYM theory studied in
\cite{BM},\cite{BH}, whereas for
$\gamma\to\infty$ one  obtains the YM-dilaton theory in
flat space  studied in \cite{LM1},\cite{Biz1}.
The value $\gamma=1$ corresponds to a model obtained from
 heterotic string theory \cite{GSW}.
It was observed \cite{LM2} that a very special
situation occurs for this value of $\gamma$.
We found strong indications that the lowest lying regular
solution may be obtained in closed form.

The rest part of this paper is organized as follows.
In the following chapter we will introduce the static
spherically symmetric ansatz and derive the corresponding
field equations. In chapter 3 we will discuss the asymptotic
behavior of solution in  the vicinity of singular points.
In chapters 4 and 5 we shall present the results of
our numerical calculations, indicating that for any value of
the dilaton coupling constant $\gamma$ the EYMD system has:

1.A discrete family of globally regular (sphaleronic type) solutions

2.A discrete family of black hole solutions.

Chapter 6 contains concluding remarks.

Our presentation here is based and closely follows Ref's
\cite{LM1},\cite{LM2}.
\vglue 0.6cm
{\bf\noindent 2. Ansatz and Field Equations}
\vglue 0.4cm
We are interested in static, spherically symmetric solutions of
EYMD theory defined by  the action Eq.~(\ref{act}).
A convenient parametrization for the metric turns out to be
\be
ds^2=A^2(r)\mu(r)dt^2-{dr^2\over\mu(r)}
  -r^2d\Omega^2\;, \label{interval}
\ee
where $d\Omega^2=d\theta^2+sin^2(\theta)d\varphi^2$
is the line element of the unit sphere.

For the $SU(2)$ YM potential we make the usual (`magnetic')
spherically symmetric ansatz
\be
W_0^a=0,\quad
  W_i^a=\epsilon_{aik}{x^k\over r^2}(W(r)-1)\;. \label{gauge}
\ee
%Note that in this parametrization $W=1$ describes the `vacuum'
%solution with the Chern-Simons number $n_{CS}=0$, while $W=-1$
%is gauge equivalent configuration with $n_{CS}=1$.

Substituting this ansatz into the action
we obtain the reduced action
\be
S_{\rm red}=-\int A\Bigl[{1\over 2G}(\mu+r\mu'-1)+
  {r^2\over 2}\mu\varphi'^2+{\e^{2\kappa\varphi}\over g^2}
  \Bigl(\mu W'^2+{(1-W^2)^2\over2r^2}\Bigr)\Bigr]\,dr\;,\label{redact}
\ee
where a prime denotes $d\over dr$.

Rescaling $\varphi\to\varphi/\sqrt G$, $r\to {\sqrt G\over g}r$ and
$S\to S/g\sqrt G$ removes the dependence on $G$ and $g$ from $S$,
hence we may put $G=g=1$ without restriction.
The only remaining parameter is the dimensionless ratio
$\gamma=\kappa/\sqrt G$, which may be also written as the ratio of
masses $\gamma={M_{Pl}\over M_D}$ with $M_{Pl}={1\over\sqrt G}$ and
$M_D={1\over\kappa}$.
The resulting field equations are

\bea
 (A\e^{2\gamma\varphi}\mu W')'&=&A\e^{2\gamma\varphi}
   {W(W^2-1)\over r^2} \;, \nn \\
 (A\mu r^2\varphi')'&=&
 2\gamma A\e^{2\gamma\varphi}\Bigl(\mu W'^2+{(1-W^2)^2\over2r^2}
 \Bigr)\;,  \nn \\
 \mu'&=&{1\over r}\Biggl(1-\mu-r^2\mu\varphi'^2-2\e^{2\gamma\varphi}
 \Bigl(\mu W'^2+{(1-W^2)^2\over2r^2}\Bigr)\Biggr)\;, \nn  \\
 A^{-1}A'&=&\frac{2\e^{2\gamma\varphi}W'^2}{r}
 +r\varphi'^2\;.\label{eqm}
\eea

A few remarks are in order.

Eq.~(\ref{eqm}) are invariant under a shift
$\varphi\to\varphi+\varphi_0$ accompanied by a simultaneous rescaling
$r\to r\e^{\gamma\varphi_0}$; hence
solutions regular at infinity can always
be normalized to $\varphi(\infty)=0$.

In the flat space limit the system of Eq.~(\ref{eqm}) reduces to
\bea
  W''&=&{W(W^2-1)\over r^2}
              -2\gamma\varphi' W'\;, \nn \\
 (r^2\varphi')'&=&
 2\gamma \e^{2\gamma\varphi}\Bigl(W'^2+{(1-W^2)^2\over2r^2}
 \Bigr)\;.\label{fleqm}
\eea

For $\gamma=0$ the dilaton field decouples and one obtains
the YM theory governed by the equation

\be
  W''={W(W^2-1)\over r^2}.
\ee
Introducing the 'time' co-ordinate $\tau=\rm{ln} r$ we obtain
 the simple equation

\be
 \ddot{W}= W(W^2-1) + \dot{W}.
\ee
where a dot denotes $d\over d\tau$.
This equation has a mechanical analogue; it describes the motion
of a 'particle' in the potential $V(W)=-(W^2-1)^2/4$ with a
negative friction due to the $\dot{W}$ term.
Globally regular solutions correspond to motions interpolating
between local maxima of the potential. Due to the friction such
motions do not exist, and hence there are no regular, static
solutiuons of the pure YM theory, in accordance with a known
result \cite{COL}.

There are some known exact solutions of Eq.(\ref{eqm}).
Their YM fields are imbeddings of abelian gauge fields,
i.e.  we are actually dealing with solutions of the
Einstein-Maxwell-dilaton theory.
All these solutions describe black holes;
no globally regular solutions of this type exist \cite{BMG}.

The simplest case is the Schwarzschild solution with trivial
YM- and dilaton fields, given by
\be
W\equiv 1,~~~~\varphi\equiv 0,~~~~A(r)=1,~~~~\mu (r)=1-
\frac{2M}{r}\; .
\ee

There is an abelian (magnetically charged) Reissner-Nordstr{\o}m
type solution with nontrivial dilaton field obtained in
\cite{DM} for special value of $\gamma=\sqrt{3}$, generalized
to arbitrary $\gamma$ in \cite{GM}.

Of particular interest is the so called 'extremal' limit of this
solution. In terms of a radial co-ordinate $\rho$ related to $r$
through $dr=r^2\sqrt{\mu}\e^{\varphi} \rho^{-2}d\rho$
this 'extremal' solution is given by expression:
\be
ds^2=\e^{2\varphi}dt^2- \e^{-2\varphi}(d\rho^2
  +\rho^2d\Omega^2)\;, \label{ext}
\ee
with
\be
\e^{2\varphi}={(1+\frac{\sqrt{\gamma^2+1}}{\rho})}^{- 2\over
{\gamma^2+1}}~~~~~{\rm and}~~~~~~ W=0.
\ee
The mass of this solution is given by
\be
 M=1/\sqrt{\gamma^2+1} \label{extmass}.
\ee

For later use we determine also the
metric coefficient $\mu$, which turns out to be given by
\be
\mu=\Biggl({1+{{\gamma^2\over\sqrt{\gamma^2+1}}\frac{1}{\rho}}
     \over{1+\frac{\sqrt{\gamma^2+1}}{\rho}}}\Biggr)^2 \label{mu}
\ee
For $\rho\to0$  the function $\mu(r)$
tends to
\be
 \mu_0=\gamma^4/(\gamma^2+1)^2 \label{mu0}.
\ee
The fact that this value differs from 1 means that the point $r=0$
is not a regular origin.

\vglue 0.6cm
{\bf\noindent 3. Asymptotic behavior of solutions}
\vglue 0.4cm
The field equations Eq.~(\ref{eqm})  have
singular points at $r=0$ and $r=\infty$ as well
as for points where $\mu(r)$ vanishes.
Regularity at $r=0$ of a configuration
requires $\mu(r)=1+O(r^2)$, $W(r)=\pm1+O(r^2)$, $\varphi(r)=\varphi(0)
+O(r^2)$ and $A(r)=A(0)+O(r^2)$.
Since $W$ and $-W$ are gauge equivalent we may choose
$W(0)=1$. Similarly we can assume $A(0)=1$ since a rescaling of $A$
corresponds to a trivial rescaling of the time coordinate.
Inserting a power series expansion into Eq.~(\ref{eqm}) one finds

\bea
W(r)&=&1-br^2+O(r^4)\;, \nn \\
\varphi(r)&=&\varphi_0+2\gamma \e^{2\gamma\varphi_0}b^2r^2
      +O(r^4)\;,\nn \\
\mu(r)&=&1-4b^2\e^{2\gamma\varphi_0}r^2+O(r^4)\;, \nn \\
A(r)&=&1+4b^2\e^{2\gamma\varphi_0}r^2+O(r^4)\;,\label{ezero}
\eea
where $b$ and $\varphi_0$ are arbitrary parameters.

Similarly assuming a power series expansion in $1\over r$ at $r=\infty$
for asymptotically flat solutions one finds
$\lim\limits_{r\to\infty}W(r)=\{\pm 1,0\}$.
It turns out that $W(\infty)=0$ cannot
occur for globally regular solutions, therefore we concentrate
on the remaining cases. One finds

\bea
 W(r)&=&\pm(1-{c\over r}+O({1\over r^2}))\;, \nn \\
 \varphi(r)&=&\varphi_\infty-{d\over r}+O({1\over r^4})\;, \nn \\
 \mu(r)&=&1-{2M\over r}+O({1\over r^2})\;, \nn \\
 A(r)&=&A_\infty(1-{d^2\over2r^2}+O({1\over r^4}))\;,\label{einfty}
\eea
where again $c,d,M,\varphi_\infty$ and $A_\infty$ are arbitrary
parameters.

Turning to singular points $r_h$, where $\mu(r)$ vanishes, we find that
solutions of Eq.~(\ref{eqm}) stay regular at such a point, if

\bea
 W(r_h+\rho)&=&W_h+W'_h\,\rho+O(\rho^2)\;, \nn \\
 \varphi(r_h+\rho)&=&\varphi_h+\varphi'_h\,\rho+O(\rho^2)\;,\nn \\
 \mu(r_h+\rho)&=&\mu'_h\,\rho+O(\rho^2)\;,  \label{bh1}
\eea

with

\bea
  W'_h&=&{W_h(W_h^2-1)\over \mu'_h r_h^2}\;,\nn \\
  \varphi'_h&=&{2\gamma\e^{2\gamma\varphi_h}(W_h^2-1)^2
        \over 2\mu'_h r_h^4}\;,\nn \\
 \mu'_h&=&{1\over r_h}\Bigl(1-{\e^{2\gamma\varphi_h}
          (W_h^2-1)^2\over r_h^2}\Bigr)\;. \label{bh2}
\eea

For a given value of $r_h$ there are the adjustable
parameters $W_h$ and $\varphi_h$ analogous to
the parameters $b$ and $\varphi_0$ at $r=0$.
If $r_h$, $W_h$ and $\varphi_h$ are chosen such that
$\mu'_h>0$ the surface $r=r_h$ describes a regular event horizon,
hence asymptotically flat solutions with this behavior
represent black holes.

The mass of such a (regular and black hole) solution is given
in units of $M_{\rm Pl}/g=\frac{1}{g\sqrt{G}}$ by
\be
M=\lim_{r\to\infty}{r(1-\mu(r))\over2}\,. \label{mass}
\ee

The generic solution with regular boundary conditions of the type
Eq.~(\ref{ezero})
or Eq.~(\ref{bh1}), Eq.~(\ref{bh2}) develops
a singularity with $\mu(r_s)=0$ at some value $r=r_s(b)$ with
finite values of $W(r_s),\varphi(r_s)$ and diverging $W'(r_s)$.
Closer analysis \cite{BFM} reveals these singularities as
coordinate singularities, where the radius $r$ is stationary.
Regular, asymptotically flat resp.\ black hole solutions
avoid this singularity and interpolate smoothly between the described
asymptotic behavior at $r=0$ resp.\ $r=r_h$ and $r=\infty$.
This may be achieved by a suitable choice of the parameter
$b$.
In fact, numerical integration of Eq.~(\ref{eqm})
indicates the existence of
discrete families of globally regular resp.\ black hole solutions
for any given $r_h$. The various members of these families are
distinguished by the number of zeros of $W(r)$.

\vglue 0.6cm
{\bf \noindent 4. Dilatonic Sphalerons \hfil}
\vglue 0.2cm
{\it\noindent 4.1. Flat Space Solution}
\vglue 0.1cm
\noindent
It is known that pure YM equations in four dimensions
have no static solutions \cite{COL}, because of the repulsive
nature of the YM self-interaction. This repulsion can be
compensated by the introduction of a Higgs field.
This way one obtains e.g.  the t'Hooft-Polyakov magnetic
monopole and the sphaleron.

The role of the attractive field can be also played by the dilaton.
In fact, it was observed \cite{LM1} (see also \cite{Biz1})
that the introduction of a dilaton field produces a discrete
sequence of particle-like solutions of finite energy.

In the next subsection we will discuss in  detail
this type of solution in presence of gravity.

\vglue 0.2cm
{\it\noindent 4.2. Gravitating Dilatonic Sphalerons}
\vglue 0.1cm
\noindent
Globally regular solutions have to interpolate
between the asymptotic behavior
Eq.~(\ref{ezero}) at $r\to 0$ and
Eq.~(\ref{einfty}) at $r\to\infty$.
Using a suitably desingularized version of Eq.~(\ref{eqm}) at $r=0$
we have integrated these equations numerically starting from $r=0$.
Due to the invariance of the field equations under a shift of
$\varphi$ accompanied by a suitable rescaling of $r$ as already
mentioned, it is sufficient to vary the parameter $b$ at the origin.
The normalization $\varphi(\infty)=0$ can be adjusted afterwards,
accompanied by a rescaling of the
parameter $b$ and of the mass $M$ of the solution.

For small values of $b$ the solution develops a zero of $\mu$ for some
$r=r_s(b)$ with $W(r_s)<-1$. On the other hand
for large $b$ the singularity occurs before $W(r)$ has a zero.
This behavior is completely analogous to the one found for the
EYM system \cite{BM}.
In that case it has been used to prove rigorously the
existence of globally regular solutions \cite{YAU1}.

For all the considered values of $\gamma$ we found a discrete set
$\{b_N\}$ for which the YM-potential $W_N$ has $N$ zeros
and then approaches $(-1)^N$.
Since with growing $N$ the precise numerical computation of
the solutions becomes increasingly difficult
we present only solutions up to $N=6$.

The numerically determined values of the parameters of the
solutions up to $N=6$ for $\gamma=1.0$ are collected in Tab 1.
The masses $M_N$ given in Tab 1.
correspond to the normalization $\varphi(\infty)=0$,
obtained by the shift $\varphi\to\varphi-\varphi_\infty$
accompanied by a rescaling $M\to M\e^{-\gamma\varphi_\infty}$.

\begin{center}
  Tab 1. Parameters of the $N=1\dots 6$ solutions
  for $\gamma =1.0$.

\vglue 0.4cm
\begin{tabular}{|l|l|l|l|l|l} \hline
 $N$ & $b$   & $\varphi_\infty$ & $M$     & $\mu_{min}$ \\ \hline
 1  &  0.166667  &  0.932284   &  0.57695 & 0.5864 \\
 2  &  0.231800  &  1.792793   &  0.68481 & 0.3705 \\
 3  &  0.246862  &  2.692205   &  0.70344 & 0.2868 \\
 4  &  0.249484  &  3.597983   &  0.70651 & 0.2637 \\
 5  &  0.249916  &  4.504705   &  0.70702 & 0.2587 \\
 6  &  0.249986  &  5.411575   &  0.70709 & 0.2534 \\ \hline
\end{tabular}
\end{center}
\vglue 0.4cm

It is remarkable that for $\gamma=1$ the parameter $b$ for the
lowest lying solution seems to be the rational number $\frac1 6$
(this is valid at least up to 12 digits).
This fact together with other regularities which we found suggest
that the $N=1$ solution for $\gamma=1$ may be obtained
in closed form.

The $\gamma$ dependence of the solutions was studied in \cite{LM2}.
We found that the mass
in natural units $M_{Pl}/g$ goes to a finite value for
$\gamma\to 0$ and drops to zero like $\frac{1}{\gamma^2}$
for $\gamma\to\infty$.

Next we would like to discuss the behavior of the discrete
family of solutions for some fixed value of $\gamma$ parametrized
by the number $N$ of zeros of $W(r)$.
Already for moderately large
$N$ the solutions develop a characteristic behavior.
Two different regions can be clearly distinguished.
%(Figs.1 and 2 illustrate this for the case $N=6$)
In an inner region the function $\mu_N(r)$ falls to the value
$\mu_0=\gamma^4/(\gamma^2+1)^2$ and then stays constant on an
interval, whose length increases with $N$.
The function $W_N(r)$ approaches a universal
form, decreasing first to $W\approx-0.15$ and then oscillating with
decaying amplitude around $W=0$. While $\mu_N$ stays more or less
constant the function $\varphi_N$ grows linearly with $\ln r$.
In a second, outer region $\mu_N$ raises
monotonically to $\mu=1$, while $W_N$ stays still very small.
Finally $W_N(r)$ raises to its asymptotic value $\pm 1$.

In the limit $N\to\infty$ we obtain two different limiting
solutions, one describing the interior part and another one the
exterior one, differing by an infinite shift of $\varphi$.
The limiting interior solution, which is regular at $r=0$,
is normalized to $\varphi(0)=0$.
It extends to arbitrarily large values of $r$.
For large $r$, where $W(r)$ is very small and $\mu(r)$ is more or
less constant, the solution can be well approximated by an
asymptotic form obtained by linearization in $W$
\bea
  W_{as}&=&r^{-{\gamma^2+1\over 2\gamma^2}}\sin{(\omega\ln
r+\delta)}\;, \nn \\
  \varphi_{as}&=&{1\over\gamma}\ln r
              -{1\over 2\gamma}\ln(\gamma^2+1)\;,\nn \\
  \mu_{as}&=&{\gamma^4\over(\gamma^2+1)^2}\;,\nn \\
  \ln A_{as}&=&A_0+{1\over\gamma^2}\ln r\;, \label{as}
\eea
with $\omega=(1+1/\gamma^2)\sqrt3/2$.

Since $\varphi_N(\infty)$  grows without limit for $N\to\infty$
the exterior limiting solution can only be obtained by renormalizing
$\varphi$ to $\varphi(\infty)=0$, accompanied by an infinite rescaling
of $r$. It turns out that $W_N(r)$ tends to zero if the rescaled value
of $r$ is kept fixed as $N$ tends to infinity.
The resulting exterior solution is the
'extremal' abelian solution discussed in chapter 2.
The mases $M_N$ of the regular solutions tend to the limiting value
 Eq.~(\ref{extmass}) from below, whereas the minima
$\mu_{min}$ tend to $\mu_0$ Eq.~(\ref{mu0}) from above.

As in  \cite{BFM}   one can find asymptotic formulae
for the $b_N$  values and for the mass $M_N$ of the solutions.
The asymptotic  formula for $b_N$ for  $\gamma=1.0$ is
\be
b_N\approx\frac{1}{4}-\frac{1}{12}e^{-{\pi\over\sqrt 3}(N-1)}\;.
\label{formula1}
\ee
A better approximation for small $N$ is given by the expression
\be
b_N\approx\frac{1}{4(1+\frac{1}{2}e^{-{\pi\over\sqrt 3}(N-1)})}\;,
\label{formula2}
\ee
where
\be
b_1={1\over 6}\;,~~ {\rm and}~~~~ b_\infty ={1\over 4} =0.25\;
\ee
have been taken into account.

Eq.~(\ref{formula1}) and  Eq.~(\ref{formula2})
are very precise already for small $N$.

All the solutions were found \cite{LM2} to be unstable
in linearized perturbation theory, as to be expected from
experience with the Bartnik-McKinnon's solutions \cite{STAB}.

\vglue 0.6cm
{\bf \noindent 5. Non-Abelian Black Holes \hfil}
\vglue 0.4cm
Black hole solutions have to interpolate
between the asymptotic behavior Eq.~(\ref{bh1}) at $r=r_h$
and Eq.~(\ref{einfty})
at $r\to\infty$.

Our numerical results indicate that there is again
a discrete infinite family of such black hole solutions for any
given values of $\gamma$ and $r_h$.
For $r_h\to 0$ the functions $W(r)$ and $\varphi(r)$ tend to
those of the regular solutions. For large values of $r_h$ the
function $\varphi(r)$ varies very little between $r_h$
and $r\to\infty$. Thus these solutions resemble those of the
EYM theory without the dilaton.
Tab. 2 contains the corresponding parameters and masses.
Renormalizing $\varphi$ to $\varphi(\infty)=0$ clearly
involves now also a rescaling of $r_h$. The values $r_{h_0}$
given in Tab. 2 are those before the rescaling (corresponding
to the normalization $\varphi(r_h)=0$).

\begin{center}
Tab 2. Size dependence of the
   black hole parameters for $\gamma=1.0$.
\vglue 0.4cm
\begin{tabular}{|l|l|l|l|l|}  \hline
$r_{h_0}$ &$r_h$  & $W_h$     & $\varphi_\infty$ &$M$    \\ \hline
0.1  &  0.040083  & 0.998331   &   0.914212      & 0.5849 \\
1.0  &  0.510138  & 0.824278   &   0.673075      & 0.6933 \\
10.0 &  9.903090  & 0.273001   &   0.009738      & 4.9997 \\
50.0 &  49.98067  & 0.268148   &   0.000387      & 24.999 \\ \hline
\end{tabular}
\end{center}
\vglue 0.4cm

Again, all these non-abelian black holes were found \cite{LM2}
to be unstable in linearized perturbation theory.

\vglue 0.6cm
{\bf \noindent 6. Concluding Remarks \hfil}
\vglue 0.4cm
We find  globally regular and black hole solutions
of a EYMD theory. The
regular solutions are sphaleron-type  configurations -
gravitating dilatonic sphalerons. One expects fermionic
zero modes and as a result unsuppressed fermion number
non-conservation processes in the background of this type of
solutions.  Due to the high mass of the
solutions  the only situation  where they could play a role
is in the Early Universe, but
 at the moment there seems to be no natural physical
scenario where we could make use of these solutions.

It is interesting that the black holes we found have
a non-vanishing YM field outside of a horizon.
As they differ from the Schwarzschild solution, but do
not carry any gauge charge we might consider them as
violating the No-Hair Conjecture (stating that black holes are
specified by their gauge charges). Yet, being unstable they do not
constitute a strong case of such a violation.

The case $\gamma=1$ corresponds to a model obtained from
string theory. We find strong indications that the $N=1$
globally regular solution can be obtained in closed form.

\vglue 0.6cm
{\bf \noindent 7. Acknowledgements \hfil}
\vglue 0.4cm
We are indebted to P. Breitenlohner and P. Forg\'acs
for many stimulating discussions.

\vglue 0.6cm
{\bf\noindent 8. References \hfil}
\vglue 0.4cm


\begin{thebibliography}{9}
\bibitem{BM}
 R. Bartnik and J. McKinnon, {\it Phys. Rev. Lett.}
{\bf 61} (1988) 141.

\bibitem{BH}
 M.S. Volkov and D.V. Gal'tsov, {\it  JETP Lett.}
{\bf 50} (1990) 346;\\
{}~~H.P. K\"unzle and A.K.M. Masood-ul-Alam, {\it J. Math. Phys.}
{\bf 31} (1990) 928;\\
{}~~P. Bizon, {\it Phys. Rev. Lett }
{\bf 61} (1990) 2844.

\bibitem{STAB}
N. Straumann and Z.H. Zhou, {\it Phys. Lett.}
{\bf B243} (1990) 33.

\bibitem{GRSPH}
 M.S. Volkov and D.V. Gal'tsov, {\it  Phys. Lett.}
{\bf B273} (1991) 255;\\
{}~~D. Sudarsky and R.M. Wald, {\it Phys. Rev.}
{\bf D46} (1992) 1453.

\bibitem{YAU1}
 J.A. Smoller, A.G. Wasserman, S.T. Yau and J.B. McLeod,
 {\it Comm. Math. Phys.} {\bf 143} (1991) 115.

\bibitem{YAU2}
 J.A. Smoller and A.G. Wasserman, {\it Comm. Math. Phys.}
{\bf 151} (1993) 303;\\
 J.A. Smoller, A.G. Wasserman and S.T. Yau, {\it Comm. Math. Phys.}
{\bf 154} (1993) 377.

\bibitem{BFM}
 P. Breitenlohner, P. Forg\'acs and D. Maison,
 {\it On Static Spherically Symmetric
 Solutions of the Einstein-Yang-Mills Equations,}
 Preprint, MPI-Ph/93-41 (1993).

\bibitem{LM1}
 G. Lavrelashvili and D. Maison, {\it Phys. Lett.}
{\bf B295} (1992) 67.

\bibitem{Biz1}
 P. Bizon, {\it Phys. Rev.}
{\bf D47} (1993) 1656.

\bibitem{LM2}
 G. Lavrelashvili and D. Maison, {\it Regular and Black Hole
 Solutions of Einstein-Yang-Mills-Dilaton Theory,}
 Preprint, MPI-Ph/92-115 (1992)\\
 to arrear in  {\it Nucl. Phys.} {\bf B}  (1993).

\bibitem{DG}
 E.E. Donets and D.V. Gal'tsov, {\it Phys. Lett.}
{\bf B302} (1993) 411.

\bibitem{Biz2}
 P. Bizon, {\it Acta Physica Polonica}
{\bf B24} (1993) 1209.

\bibitem{MAEDA}
 T. Torii and K. Maeda, {\it Black Holes with
 Non-Abelian Hair and their Thermodinamical Properties,}
Waseda University preprint, WU-AP/28/93.

\bibitem{GSW}
M.B. Green, J.H. Schwarz and E. Witten
{\it Superstring Theory,} Cambridge U.P., Cambridge, 1987.

\bibitem{COL}
 S. Coleman, in: {\it New Phenomena in Subnuclear Physics,}
ed. A. Zichichi, Plenum, NY, 1975;\\
 S. Deser, {\it Phys. Lett.} {\bf B64} (1976) 463.

\bibitem{BMG}
 P. Breitenlohner, D. Maison and G. Gibbons,
 {\it Comm. Math. Phys.} {\bf 120} (1988) 295.

\bibitem{DM}
 P. Dobiasch and D. Maison, {\it Gen. Rel. and Grav.}
{\bf 14} (1982) 231.

\bibitem{GM}
 G.W. Gibbons and K. Maeda, {\it Nucl. Phys.}
{\bf B298} (1988) 741.


\end{thebibliography}
\end{document}